\documentclass{moriond}

\def\be{\begin{equation}}
\def\ee{\end{equation}}
\def\bea{\begin{eqnarray}}
\def\eea{\end{eqnarray}}

\usepackage{wrapfig}
\usepackage{amsmath,amsfonts}

\begin{document}
\vspace*{4cm}
\title{Assessing the sensitivity to Axion-Like-Particle Dark Matter with very-high-energy gamma-ray observations of selected AGN and galaxy cluster pairs}

\author{C. Grimaud$^1$, D. Malyshev$^2$, E. Moulin$^1$}

\address{$^1$Irfu, CEA Saclay, Universit\'e Paris-Saclay, F-91191 Gif-sur-Yvette, France \\
$^2$Institut f\"ur Astronomie und Astrophysik, Universit\"at T\"ubingen, Sand 1, D 72076 T\"ubingen, Germany}

\maketitle\abstracts{
Axion-Like-Particles (ALPs) are pseudo-scalar particles actively searched as light dark matter candidates. ALPs can couple to photons which give rise to the possibility of oscillations with photons in an external magnetic field. If sufficiently strong, this coupling can imprint distinctive spectral irregularities in the gamma ray spectrum of astrophysical sources.
We present a prospective study on the sensitivity of probing ALP-photon interactions using stacked observations of selected active galactic nuclei (AGNs) located behind galaxy clusters. The ALP-photon conversion in cluster magnetic fields produces absorption-like features in AGN spectra that are difficult to predict for individual sources. To address this, we apply a stacking analysis of multiple AGN-cluster pairs, yielding a controlled prediction of the expected ALP induced spectral patterns and enhancing the sensitivity to such irregularities.
Using simulated data for selected hard-spectrum Fermi/LAT AGNs that can be observed by Imaging Atmospheric Cherenkov Telescopes such as H.E.S.S., we evaluate the performance of this method. The combination of mock IACT observations with our stacking approach enable exploration of the previously uncharted ALP dark matter parameter space in the neV mass range.}

\section{Introduction}
Ever since dark matter was theorized to explain the gravitational effects witnessed over a wide range of astrophysical and cosmological observations, the search for it has been one of the main challenges of modern physics. Among the compelling dark matter candidates, this article will focus on axion-like particles (ALPs). ALP are pseudo-Nambu-Goldstone bosons resulting from the breaking of a U(1) symmetry and are a generalization of the QCD axion postulated to solve the strong CP problem. Contrary to the axion, the ALP coupling is independent of its mass resulting in a larger parameter space to explore. 
A property of ALPs is their two-photon vertex which in the presence of an external magnetic field can lead to ALP-photon oscillations. This is especially interesting in the context of very high energy gamma ray observations with the current and future network of Imaging Atmospheric Cherenkov Telescopes since ALP-photon oscillations can be probed through its effects on VHE gamma rays spectra.
The probability of ALP-photon oscillation is dependent on the external magnetic field strength value and on the length of the ALP propagation through the magnetized region. In this article we will focus on VHE gamma ray emitted by Active Galactic Nuclei (AGN) and probing the ALP-photon oscillations in galaxy cluster magnetic fields. This offers the advantage of existing estimation of the magnetic field strength~\cite{Govoni:2004as,Bonafede:2010xg}
but has the disadvantage of lacking reliable measurements of the small-scale properties of the field such as orientation of the field in domains and measurement of the domains lengths.
To overcome this issue, a stacking analysis of multiple AGN-galaxy cluster pairs has been proposed~\cite{Malyshev:2025iis}. This allows to obtain a smooth and predictable spectral imprint to probe ALP-photon oscillations. This technique leads to an enhancement of the statistical power and allows to derive a robust constraint on ALP parameters. In this article we present the results of this stacking analysis method using mock data simulated for H.E.S.S., MAGIC and VERITAS observations of selected AGN-galaxy cluster pairs.

\section{Selection of AGN-GC pairs for mock observations}

\begin{wrapfigure}{r}{0.5\textwidth}
  \begin{center}
    \includegraphics[width=0.48\textwidth]{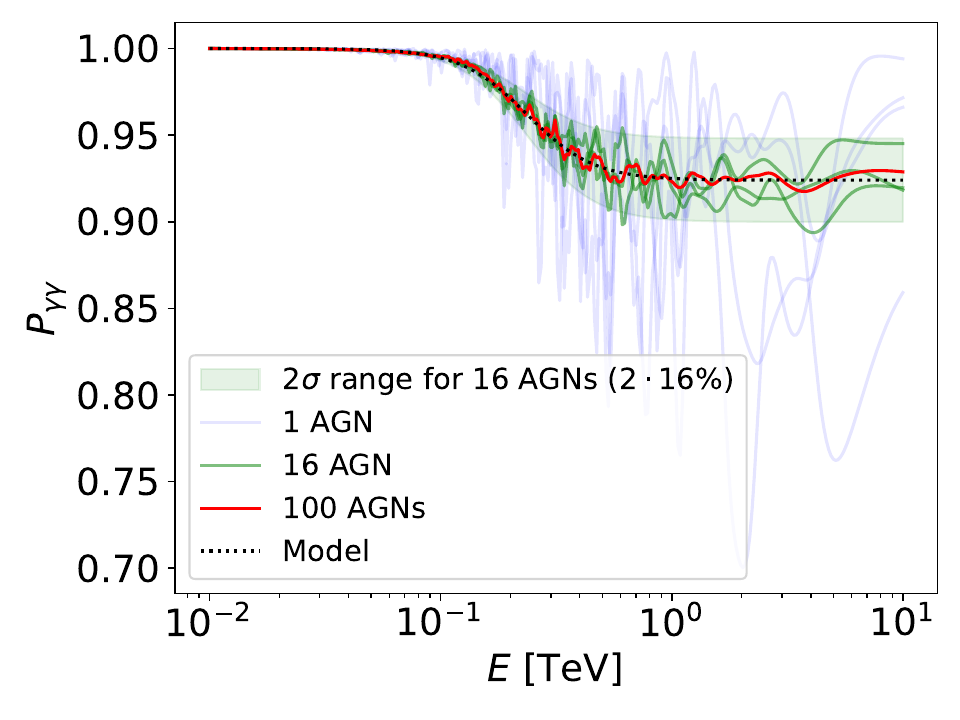}
  \end{center}
  \caption{\label{fig:pgammagamma}Photon survival probability $P_{\gamma\gamma} $and its averages as a function on the photon energy. The blue lines are the probability for different realizations of the cluster magnetic field properties. All realizations have the same radial profile of the magnetic field but vary randomly in their orientation in the photon polarization plane and in coherence length of the magnetic domains. The green and red lines show the effect of averaging over 16 and 100 randomly selected realizations, respectively. Black dotted-dashed line shows the analytical approximation to these lines in case of infinite number of objects.  
  The ALP parameters for all curves are $(m_a, g_{a\gamma\gamma})$ = (20 neV, 7$\times$10$^{-13}$ GeV$^{-1}$).}
\end{wrapfigure}

The AGN-galaxy cluster pairs selection was made by cross-matching the high-latitude Fourth Catalog of Active Galactic Nuclei 4LAC-DR3-h~\cite{4lacdr3} of gamma-ray bright AGNs detected by Fermi-LAT with the Sunaev-Zeldovich catalogues of optically and X-ray identified clusters of galaxies~\cite{2018MNRAS.475..343W,planck_2015sz,erass1_clusters}. Within these catalogues, we selected AGNs and clusters pairs for which the redshift of the AGN is larger than the one of the galaxy cluster and the line of sight of the AGN is within 500~kpc comoving distance from the galaxy cluster's center. In addition, from this selection, we kept only the AGNs for which the best fit given by the Fermi-LAT catalogue was a power law. This allowed to select 29 pairs from which we kept only the 16 AGN also present in the 3FHL catalog~\cite{Fermi-LAT:2017sxy} of sources detected by Fermi-LAT above 10~GeV. Finally we restricted the pairs to objects visible by currently operating IACTs (H.E.S.S., MAGIC and VERITAS) at more than 30$^{\circ}$ above horizon meaning for which the minimal zenith angle is inferior to 60$^{\circ}$. 
Following this criteria, we selected 11 AGN-cluster pairs suitable for this study for H.E.S.S. and 15 pairs suitable for both MAGIC and VERITAS. In total, this leads to 41 independent observation mock datasets that can be used to study the sensitivity to ALP.

\section{Expected VHE gamma-ray signals}

When a VHE gamma ray goes through a galaxy cluster magnetic field it has a given probability of interacting with an ALP through ALP-photon oscillation. In this study we focus on cool-core clusters, where the magnetic field is commonly modeled as scaling with the electron density and reaching $\sim 10,\mu$G in central regions. The thermal electron density is described by a $\beta$-model, $n_e(r) = n_0 \left(1 + \frac{r^2}{r_c^2} \right)^{-3\beta/2}$, with $n_0 = 3.44 \times 10^{-3} , cm^{-3}$, $r_c = 291 ,kpc$, and $\beta = 0.75$~\cite{Bonafede:2010xg,2003MNRAS.343..401L}, and the magnetic field follows $B(r) = B_0 \left( \frac{n_e(r)}{n_0} \right)^{0.67}$, vanishing beyond half the propagation distance. In a galaxy cluster magnetic field the VHE gamma ray survival probability $P{\gamma\gamma}$ is dependent on the realization of the magnetic field (orientation and coherence length in the magnetized object) as presented with the blue lines in figure \ref{fig:pgammagamma}.

This means that to use only one pair of AGN-cluster, we would need to marginalize over a many realizations of the magnetic field which would lead to a weaker constraint. Stacking several pairs allow to smooth the expected oscillation pattern and parametrize the photon survival probability as a step function $\langle P_{\gamma \gamma} \rangle = 1 - p_0/(1 + (E/E_c)^k)$ where $p_0$ is the suppression depth at high energies, $E_c$ is the transition energy and $k$ the sharpness of the step These parameters are function of the ALP parameters $(m_a,g_{a\gamma\gamma})$. This corresponds to the green and red lines in Figure~\ref{fig:pgammagamma}.

The differential gamma-ray flux expected to be observed on Earth from an AGN with a GC along the line of sight can then be written as
\begin{equation}
\frac{d\Phi^{\rm obs}_\gamma (E_{\gamma})}{dE_{\gamma}}= \frac{d\Phi^{\rm int}_\gamma (E_{\gamma})}{dE_{\gamma}} \times P_{\gamma\gamma}(E_\gamma, m_a, B_T) \times e^{-\tau(E_\gamma,\epsilon,z)}\, ,
    \label{eq:flux}  
\end{equation}
where $e^{-\tau(E_\gamma,\epsilon,z)}$ corresponds to the extragalactic background light (EBL) absorption effects.
Then to simulate our mock data, the model-expected number of events per $i^{th}$ bin with a width $\Delta E_i$ and for the $k^{th}$ pair can expressed as
\begin{equation}
    N^\mathcal{M}_{ik} (m_a, g_{a\gamma\gamma}, \mathcal{P}_k) = T_{{\rm obs}, k} \int\limits_{E_i - \Delta E_i /2}^{E_i + \Delta E_i /2} dE  \int\limits_{-\infty}^{\infty} dE'\, \frac{d\Phi^\mathcal{M}_{k}(E',\mathcal{P}_k)}{dE'}\, 
    A_{{\rm eff}, k}^{\gamma}(E')\, G(E - E')\, , 
\end{equation}
where $E$ and $E'$ are the true and reconstructed energies, $T_{obs,k}$ is the observation time of the pair $k$, $A_{eff, k}^\gamma$ is the energy-dependent effective area of the detector and $G$ is the energy resolution of the detector. In this study, we simulated data for $T_{obs} = 50$ h for each of the $k$ AGN-GC pairs and for energies from 100~GeV to 60~TeV.

\section{Data analysis of combined AGN observations with IACTs}
The statistical analysis and the computation of the expected sensitivity for a given IACT are done following the definition of a log-likelihood ratio test statistic (TS). We use binned Poisson likelihood for the observed counts $N^{\mathcal{O}}_{ik}$ and the expected counts $N^{\mathcal{M}}_{ik}$ defined as $log \mathcal{L}_{ik} = N^{\mathcal{O}}_{ik} \;\, log N^{\mathcal{M}}_{ik} - N^{\mathcal{M}}_{ik} - log (N^{\mathcal{O}}_{ik} !)$.
We compare our mock spectra under two hypotheses, the null hypotheses for which there is no ALP-photon oscillation and the ALP hypotheses for which the spectra are modulated due to ALP-photon oscillations.
Finally, the test statistic is defined as the minimization of the loglikelihood ratio of our two hypotheses as 
\begin{equation}
    TS(m_a,g_{a\gamma\gamma}) = \underset{P^1,P^2}{min} \left[ -2 log \; \frac{\underset{i,k}{\prod} \mathcal{L} ( N^{\mathcal{O}}_{ik}(m_a,g_{a\gamma\gamma},P^1),N^{\mathcal{M}}_{ik}) }{\underset{i,k}{\prod}\mathcal{L} ( N^{\mathcal{O}}_{ik}(m_a,g_{a\gamma\gamma=0},P^2),N^{\mathcal{M}}_{ik})} \right] ,
\end{equation}
where $P^1$ and $P^2$ are the parameters of the model with ALP and the model without ALP, respectively. Then, the one sided 95\% C.L. upper limit can be defined by solving for the ALP-photon coupling above the best fit where TS = 2.71.

\section{Results}

\begin{wrapfigure}{r}{0.5\textwidth}
  \begin{center}
    \includegraphics[width=0.48\textwidth]{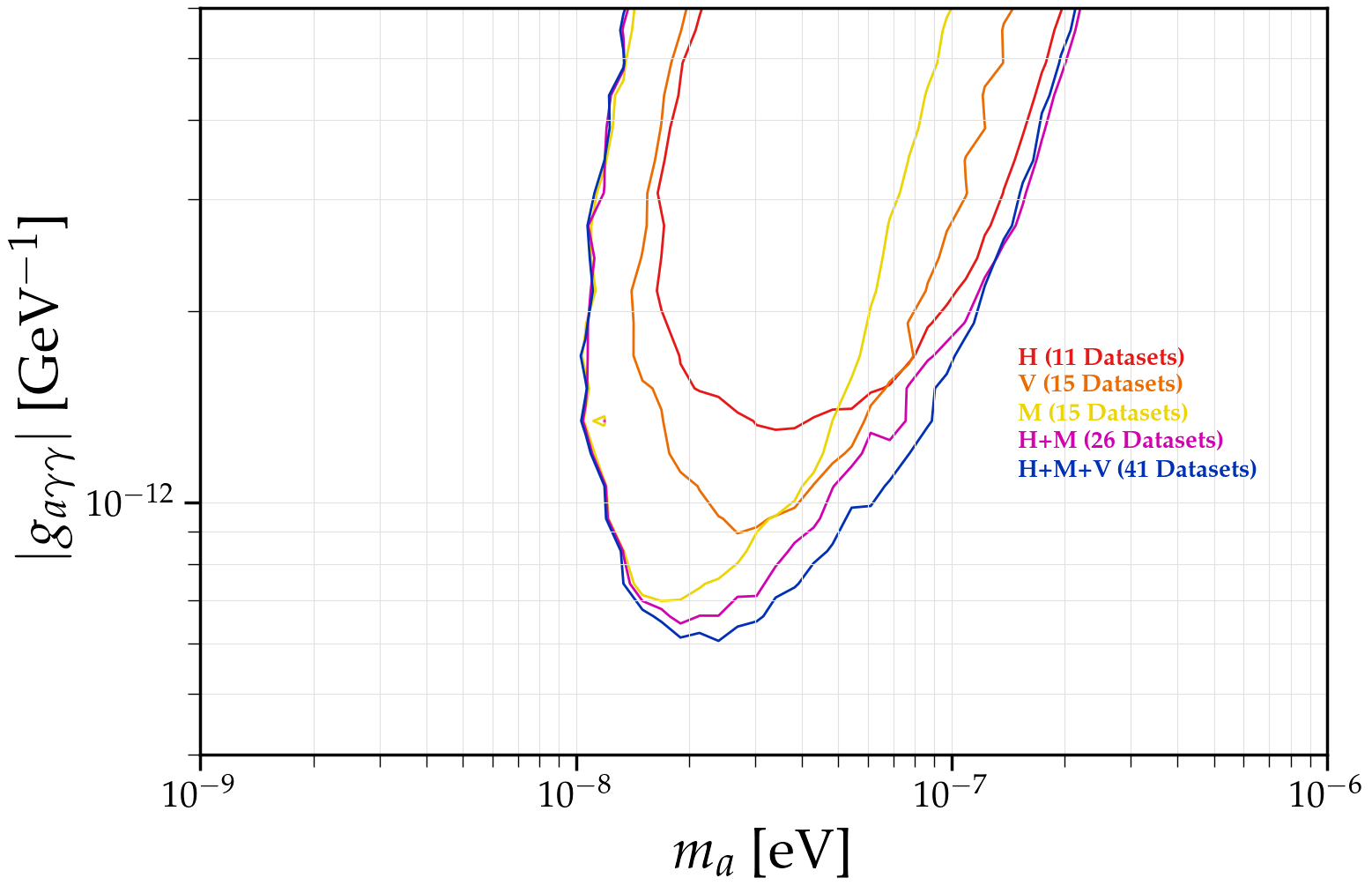}
  \end{center}
  \caption{Sensitivity on the $g_{a\gamma\gamma}$ coupling versus ALP mass $m_a$  for combined mock datasets of IACT observations assuming the Francheschini EBL model. The sensitivity is expressed as the 95\% C. L. mean expected upper limit. The sensitivity is given for mock observations of 11 AGNs by H.E.S.S. (H, red line), 15 by VERITAS (V, orange line), 15 AGNs by MAGIC (M, yellow line). Combined mock observations by H.E.S.S. and  MAGIC (H+V, 26 datasets, pink line), and by H.E.S.S., MAGIC and VERITAS (H+M+V, 41 datasets, blue line), respectively, are also displayed.}
    \label{fig:res_alldat_ebl}
\end{wrapfigure}

The plot in the left panel of Figure~\ref{fig:res_alldat_ebl} demonstrates how combining more datasets, from 11 with H.E.S.S. up to 41 for H.E.S.S., MAGIC, and VERITAS combined, sharpens the sensitivity to ALPs. At an ALP mass of m$_a$ = 3$\times$10$^{-8}$ eV, the coupling limit reaches g$_{a\gamma\gamma}$ = 6$\times$10$^{-13}$ GeV$^{-1}$, which corresponds to a factor three improvement from 11 to 41 datasets.
The sensitivity drops near 10$^{-8}$ eV as the critical energy approaches the IACTs low-energy threshold, and further weakens at a few 10$^{-7}$ eV due to strong EBL absorption and reduced photon statistics.

To assess the impact of EBL model selection, we tested several models Dominguez \textit{et al.}\cite{Dominguez:2010bv}, Franceschini \textit{et al.}\cite{Franceschini:2017iwq}, and Finke \textit{et al.}\cite{Finke:2009xi}. We found that the choice of EBL model affects sensitivity by up to 3\% for $m_a < 7\times10^{-8}$ eV, rising to 6\% for $m_a \leq 2\times10^{-7}$ eV in the 41 dataset analysis.

In addition, we tested the effect of EBL mismodeling by using mock data generated with the Franceschini EBL model but analyzed with the Finke model. We were able to reveal a false ALP signal detection (TS $\sim$ -0.50) when using the 11 H.E.S.S. datasets, driven by the different energy-dependencies of the EBL attenuation in Finke and Franceschini models below 10 TeV for low-redshift sources ($z < 0.2$). This effect fades with 41 datasets, as higher-redshift sources reduce this bias.
We also showed that a statistically homogeneous dataset, avoiding over-representation of bright-nearby objects, minimizes EBL mismodelling effects and ensures robust ALP searches, see~\cite{grimaud2026} for more details.

\begin{wrapfigure}{r}{0.55\textwidth}
  \begin{center}
    \includegraphics[width=0.53\textwidth]{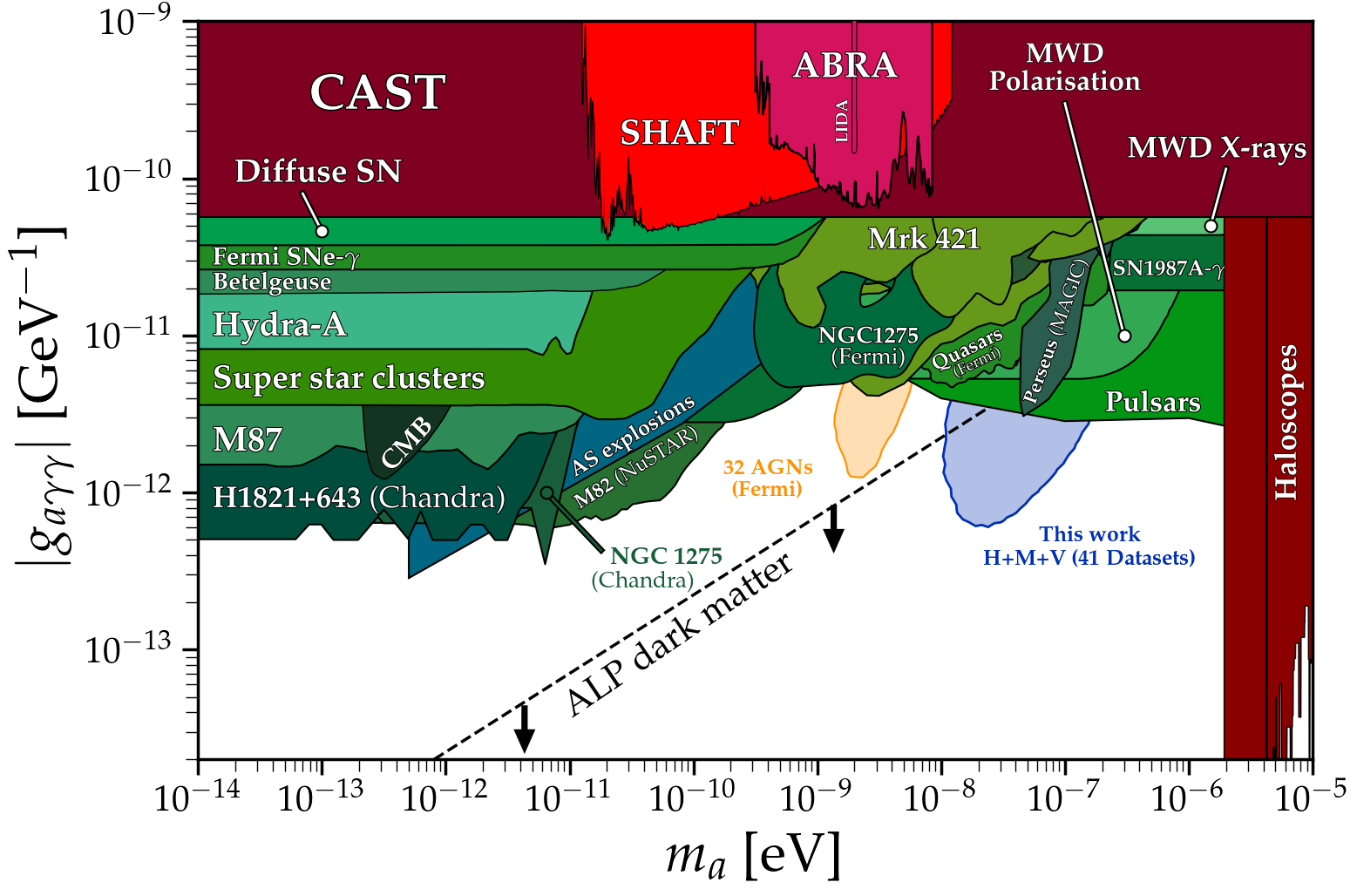}
  \end{center}
  \caption{Sensitivity on the ALP-photon coupling $g_{a\gamma\gamma}$ versus the ALP mass $m_a$. The sensitivity is expressed as 95\% C.L. mean expected upper limits. The sensitivity is  derived from combined mock observations by H.E.S.S., MAGIC and VERITAS, resulting in 41 independent datasets (solid blue line).  Also shown are the  exclusion regions from current experiment. Below the dashed thick line corresponds to the region of the parameter space where ALP can comprise all the DM in the universe. 95\% C.L. upper limits derived from the stacked analysis of observations of 32 AGN-GC pairs with Fermi-LAT and studied are shown as the solid orange line.}
    \label{fig:summaryplot}
\end{wrapfigure}

Finally, Figure~\ref{fig:summaryplot} compares the projected sensitivity of combined H.E.S.S., MAGIC, and VERITAS mock observations with existing limits from experiments (CAST, SHAFT), astrophysical searches (Chandra, Fermi-LAT, IACTs), pulsar radio data, and haloscopes~\cite{axionlimits}.

\section{Conclusions}
This study presents a new analysis strategy using current IACTs like H.E.S.S., MAGIC, and VERITAS to search for ALPs by observing VHE gamma ray from AGNs behind galaxy clusters. By combining mock observations of 41 AGNs, we show that this approach can explore yet uncharted ALP parameter space in the 10 to 100 neV mass range and couplings above $5\times 10^{-13}$~GeV$^{-1}$, a range where ALPs could potentially account for all dark matter in the Universe.

Although the method requires large observation times (550 hours for H.E.S.S., and 750 hours for MAGIC and VERITAS), it is feasible over several years. While statistical uncertainties dominate, systematic effects (such as EBL and magnetic field modellings) can be mitigated with future multiwavelength observations and improved simulations. The study concludes that this stacking analysis is a promising way to probe ALP dark matter, especially with future observatories like the CTAO.

\section*{References}
\bibliography{moriond}

\end{document}